\documentclass[10pt,conference]{IEEEtran}
\IEEEoverridecommandlockouts
\usepackage{cite}
\usepackage{amsmath,amssymb,amsfonts}
\usepackage{algorithmic}
\usepackage{graphicx}
\usepackage{textcomp}
\usepackage{xcolor}
\usepackage{amsthm}
\usepackage{subfigure}
\usepackage{colortbl}
\usepackage{soul}  
\usepackage{enumitem}
\usepackage{listings}
\usepackage{xcolor}
\usepackage{fancyhdr}
\usepackage{makecell}
\usepackage{soul}
\definecolor{skyblue}{rgb}{0., 0.72, 0.92}
\usepackage{amsmath, amsfonts}
\usepackage[outline]{contour}
\usepackage[linesnumbered,ruled,vlined]{algorithm2e}
\SetKwInOut{Input}{Input}
\SetKwInOut{Output}{Output}

\SetCommentSty{mycommfont}

\usepackage{multirow}
\usepackage{booktabs}
\usepackage{balance}

\usepackage[utf8]{inputenc} 
\usepackage[T1]{fontenc}    
\usepackage{hyperref}       
\usepackage{url}            
\usepackage{booktabs}       
\usepackage{amsfonts}       
\usepackage{nicefrac}       
\usepackage{microtype}      
\usepackage{multirow}
\usepackage{rotating}
\usepackage{bbding}
\usepackage{subfigure}
\usepackage{amssymb}
\usepackage{listings}
\usepackage{enumitem}
\hypersetup{
    colorlinks=true,
    linkcolor=blue,
    filecolor=magenta,      
    urlcolor=blue,
    pdftitle={Overleaf Example},
    pdfpagemode=FullScreen,
    }
\usepackage{amsmath}

\theoremstyle{definition}

\definecolor{lightgray}{rgb}{0.78, 0.78, 0.78}
\sethlcolor{lightgray}  
\definecolor{answercolor}{RGB}{240, 240, 240}

\def\BibTeX{{\rm B\kern-.05em{\sc i\kern-.025em b}\kern-.08em
    T\kern-.1667em\lower.7ex\hbox{E}\kern-.125emX}}
\begin{document}

\title{Large Language Model Supply Chain: Open Problems From the Security Perspective}

\DeclareRobustCommand{\IEEEauthorrefmark}[1]{\smash{\textsuperscript{\footnotesize #1}}}

\author{\IEEEauthorblockN{Qiang Hu\IEEEauthorrefmark{1},
Xiaofei Xie\IEEEauthorrefmark{2}, 
Sen Chen\IEEEauthorrefmark{3},
and Lei Ma\IEEEauthorrefmark{1,}\IEEEauthorrefmark{4}}
\IEEEauthorblockA{\IEEEauthorrefmark{1}The University of Tokyo, Japan\\
\IEEEauthorrefmark{2}Singapore Management University, Singapore  \\
\IEEEauthorrefmark{3}Tianjin University, China  \\
\IEEEauthorrefmark{4}University of Alberta, Canada
}}


\maketitle
\thispagestyle{fancy}
\pagestyle{fancy}
\cfoot{\thepage}
\renewcommand{\headrulewidth}{0pt} 
\renewcommand{\footrulewidth}{0pt}

\begin{abstract}

Large Language Model~(LLM) is changing the software development paradigm and has gained huge attention from both academia and industry. Researchers and developers collaboratively explore how to leverage the powerful problem-solving ability of LLMs for specific domain tasks. Due to the wide usage of LLM-based applications, e.g., ChatGPT, multiple works have been proposed to ensure the security of LLM systems.
However, a comprehensive understanding of the entire processes of LLM system construction~(the LLM supply chain) is crucial but relevant works are limited. More importantly, the security issues hidden in the LLM SC which could highly impact the reliable usage of LLMs are lack of exploration. Existing works mainly focus on assuring the quality of LLM from the model level, security assurance for the entire LLM SC is ignored. In this work, we take the first step to discuss the potential security risks in each component as well as the integration between components of LLM SC. We summarize 12 security-related risks and provide promising guidance to help build safer LLM systems. We hope our work can facilitate the evolution of artificial general intelligence with secure LLM ecosystems.


\end{abstract}


\section{Introduction}
\label{sec:intro}

Large Language Models~(LLMs) lead the new era of artificial intelligence~(AI). With this success, LLM-driven applications have achieved exciting and even human-better results in multiple domains, such as video generation~\cite{li2024llava}, mathematical competition~\cite{mao-etal-2024-champ}, code generation~\cite{liu2024your}, and autonomous driving~\cite{xu2024drivegpt4}. Recently, many big-tech companies have been trying to develop their own LLMs and construct relevant systems for specific tasks and products, pushing researchers and developers to explore more reliable LLM system construction roads.

Similar to conventional software systems, the construction of LLM systems consists of multiple components and participators, e.g., data providers and model developers. The integration of those complex components is called supply chain~(SC) from the perspective of software engineering where software~(LLM) serves as the core surrounding numerous upstream and downstream participants. More importantly, as LLMs are widely used nowadays, and sometimes in safety and security-critical situations such as autonomous driving systems, ensuring the reliability of the whole LLM system becomes important.
While the supply chain of conventional software systems has been extensively studied over the past decade, research specifically focused on the supply chain of AI systems, particularly LLMs, remains limited.

Although recent studies explore security-related issues in LLMs, most of them remain focused on the model itself. We analyzed 59 relevant papers from recent surveys~\cite{wang2024large, huang2024survey, fan2023trustworthiness, chang2024survey}, and found that 19 of these still concentrate on model-level security, while 24 are exclusively focused on ChatGPT, largely overlooking the broader supply chain. For instance, researchers have shown that adversarial attack techniques can be used to exploit LLMs~\cite{dong2023robust}, and LLMs can be easily jailbroken~\cite{deng2023jailbreaker}. However, the model is just one component of an LLM system. Even if the model security is ensured, vulnerabilities in other parts of the LLM supply chain, such as third-party dependencies or deployment environments, can still pose significant security risks and lead to an unreliable LLM system. Therefore, there is a critical need for research on LLM supply chain security. 

While the recent study~\cite{wang2024large} explores components of the LLM supply chain, including infrastructure, model lifecycle, and the application ecosystem, its focus remains primarily on the model lifecycle, considering each component separately. The dependencies (i.e., upstream and downstream components) within the LLM supply chain and the associated security risks along the supply chain are still unclear.

\begin{figure*}[ht]
    \centering
    
    \includegraphics[scale=0.4]{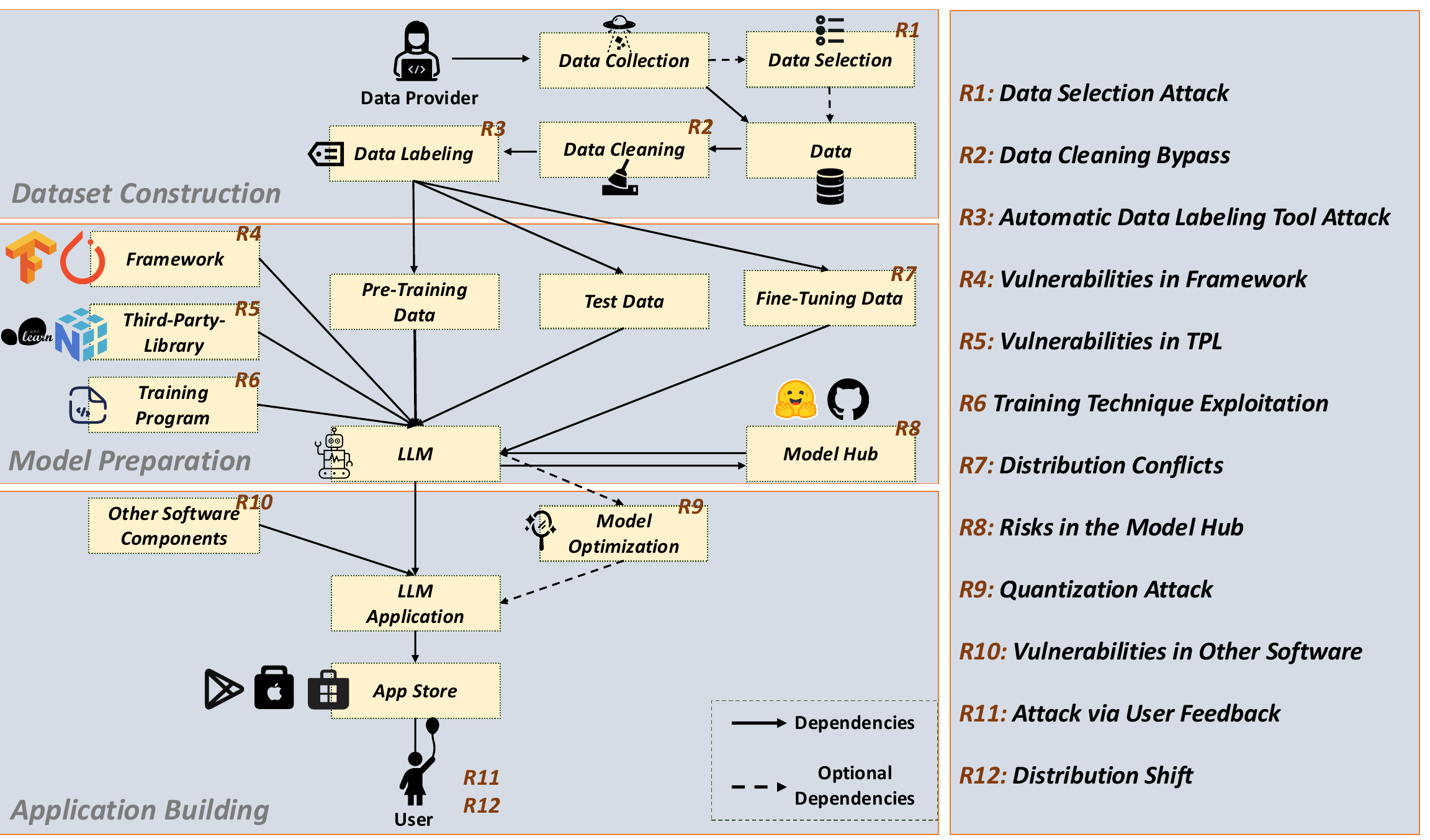}
    \label{fig:overview}
    \caption{Security risks in LLM supply chain.}
\end{figure*}

To fill this gap, we take the first step in analyzing potential security risks in the LLM supply chain and propose guidelines to mitigate these risks.
Specifically, we begin by defining each component of the LLM supply chain, analyzing dependencies of those components, e.g., from upstream data providers to downstream LLM applications (end users). We then focus on identifying security risks that originate from upstream components, which attackers could exploit to impact downstream participants. For instance, attackers may inject poisoned data into training datasets (upstream), affecting the trained LLMs and, consequently, the deployed LLM-based applications (downstream). In total, we identify 12 security risks related to the LLM supply chain and offer guidance to support the construction of secure LLM systems. We believe that our work can help researchers and developers better understand LLM supply chain security and build more reliable LLM systems.

To summarize, the main contributions of this paper are:

\begin{itemize}[leftmargin=*]
\itemsep0em
\item We are the first to explore security risks considering the integration of dependent components in the LLM SC where we summarize 12 relevant LLM SC risks.
\item We provide promising guidelines to mitigate the risks in developing LLM systems.
\end{itemize}

\section{Risks in Large Language Supply Chain}
\label{sec:llm sc}


\subsection{LLM Supply Chain Extraction}
To understand the potential security risks, we first identify the components in the LLM supply chain. Each component could affect not only itself but also the downstream participants. The process of identifying the LLM supply chain involves tracing the potential upstream components that influence or depend on a given component. Specifically, we start with downstream users who may seek an LLM application from an App store or the Internet. The App store then becomes a potential upstream component. The LLM application in the store has further upstream components, such as the LLM and other software dependencies. By iteratively tracing these dependencies, we continue identifying relevant components until we reach the highest upstream elements where we choose to stop based on the relevance. Finally, we constructed a dependency graph representing the LLM supply chain, as shown in Figure~\ref{fig:overview}.

\subsection{LLM Supply Chain}

Based on our extracted LLM supply chain, we introduce each component in the chain from the data provider to the end user~(from upstream to downstream). In the data preparation phase, data analysts collect data from the data providers and do data cleaning to filter noisy data and features. After that, data labelers~(human or labeling machine) assign ground truth to the raw data which will be utilized in the following training and testing processes. Due to the large amount of data we can obtain daily, data selection is an optional process to reduce the data processing effort. In this data construction phase, the reliability of data selection, data cleaning, and automatic data labeling tools highly affects all the remaining downstream components.

The downstream phase of data preparation is model construction. Roughly speaking, there are three ways to prepare a LLM, 1) prepare it from scratch, 2) download it from the open-source model hubs and directly use it, and 3) download it and then fine-tune it for specific domain tasks. When preparing LLMs from scratch, developers utilize AI frameworks and third-party libraries to write the training program, and then train the model with pre-training data provided by the data analysts. For the second way, developers directly employ the downloaded LLMs for the target task. For the last way, the downloaded LLMs are further fine-tuned using fine-tuning datasets to fit new data distribution.  After the model training, test reports are generated based on the test dataset, which will be provided to the application developers along with the trained LLMs. In this model construction phase, concerns could arise due to the vulnerabilities in the program, the model training process, the conflicts between pre-training and fine-tuning, and malicious hidden in model hubs. 

After application developers obtain the model, based on the specific requirements such as the application size limitation, they will decide whether to utilize model optimization techniques to reduce the model size or not. Next, LLMs are integrated with other software, including additional AI models, to form the final LLM application package, which is then uploaded to the application store. Finally, the downstream users download the application for real-world usage. In this application preparation phase, risks are potentially hidden in other software and the model optimization process.


\subsection{LLM Supply Chain Risk Analysis}
\label{sec:llm_sc_security_analysis}
Based on the LLM supply chain, we discuss potential security risks by identifying attack paths that originate from upstream components and target downstream users. These attack paths allow adversaries to compromise users by executing malicious operations at upstream points. In total, we identify and summarize 12 potential risks across the entire LLM SC.

First, as data is the key driver for LLM, security assessment of datasets is important where training data will affect the whole LLM SC. Treating the data provider as the upstream supplier, the downstream participant should be the data analyst. Existing works~\cite{chang2024survey} already summarized the risks in the training data from different perspectives, such as backdoor attacks, poisoning attacks, privacy leakage, unauthorized disclosure and privacy concerns, bias, toxicity, and ethics. 
Some risks directly impact downstream components, such as poisoning attacks that degrade the performance of trained LLMs and inject backdoors, compromising model predictions and user security. From the perspective of the LLM supply chain, several security risks need to be addressed:

\ul{\textit{Risk1: Risks in data selection process}}. Utilizing all of them to build LLM systems is almost impossible due to the large amount of available data. Data selection is a technique of filtering high-quality data for training or testing and will affect all the remaining components in the LLM SC. Recent research~\cite{sun2023robust, hu2023evaluating} demonstrated that existing automatic data selection methods have limitations such as only considering uncertain data which can be exploited by attackers to inject malicious into the collected dataset. For example, attackers can inject backdoors into the data that have high uncertainty scores to multiple common LLMs, and lead the data selection methods to select certain data to the dataset.

\ul{\textit{Risk2: Risks in data cleaning process}}. After collecting datasets, data analysts will employ data cleaning techniques to remove redundant and noisy data~(and features) to increase the quality of collected datasets. However, potential risks are hidden in this process, 1) existing data cleaning techniques are relatively simple, such as rule-based error sample detection~\cite{miao2021rotom}. It is easy to bypass such techniques by constraint-ware adversarial attacks~\cite{simonetto2021unified} and introduce malicious to the datasets. 2) More generally, even though data scientists can guarantee the security of cleaned data for the downstream model training by collaboration with model developers, it is unclear whether the cleaned data still contains risks that affect further LLM SC components such as model maintenance and model compression.     

\ul{\textit{Risk3: Risks in data labeling process}}. Normally, given a domain task, a useful data sample requires the raw data and the corresponding ground truth. Labeling a large amount of raw data is challenging and may naturally introduce label errors. For example, even for simple image classification tasks, incorrect labels exist in the dataset after carefully manually labeling~\cite{northcutt2021pervasive}. This mislabeling issue can mislead the model training and testing, and harm the performance of LLMs. Besides, multiple automatic data labeling techniques and frameworks have been proposed recently~\cite{wang2021want}. The other risk is such tools also rely on deep learning models and can be fooled to generate incorrect labels by adversarial attacks.

In addition to the risks in the data, there are also many security issues hidden in the model preparation phase. For example, the vulnerabilities in the training program, the improper use of training methods, and the different distribution between the pre-training and fine-tuning datasets can affect the produced LLMs and further introduce different reliability issues to the downstream components.

\ul{\textit{Risk4 and Risk5: Vulnerabilities hidden in the AI framework and third-party-libraries}}.  Even though many works have attempted to reveal the vulnerabilities in the DL frameworks and other third-party libraries to ensure safe LLM development, they mainly focus on the influence of such vulnerabilities on the software implementation and the AI models. Even worse, some vulnerabilities could arise in later downstream components in the LLM SC, for example, existing works~\cite{guo2019empirical} found that implementation errors in the framework can introduce serious prediction errors of AI applications on the website and cause unreliable outputs. Therefore, in addition to the software and models, analyzing the potential influence of vulnerabilities on the following LLM SC is necessary. Besides, recent works demonstrated that attackers can modify APIs in the framework and inject backdoors into the binary model files to assess information from users~\cite{modelscan}. Risk assurance of frameworks and TPLs is crucial for securing the usage of LLM applications for users.

\ul{\textit{Risk6: Risks introduced in training techniques}}. Researchers proposed multiple training techniques such as active learning and adversarial training to prepare models with different purposes. Recent works reveal that such training techniques will introduce different reliability issues to the compressed models, e.g., active learning reduces the adversarial robustness of models after quantization~\cite{hu2021towards}. Therefore, it is needed to understand how different training techniques affect the following SC components. Besides, training techniques can be also exploited by attackers. For instance, similar to the issues mentioned in the data selection process, it is possible to attack active learning methods to force models to select vulnerable data for training. 

\ul{\textit{Risk7: Distribution conflicts between pertaining datasets and fine-tuning datasets}}. As shown in Figure~\ref{fig:overview}, LLMs can be prepared by fine-tuning the pre-trained model given specific domain datasets. Unfortunately, as shown in some recent works~\cite{li2017learning, hu2022empirical}, pertaining datasets and fine-tuning datasets can sometimes have distribution conflicts and lead to the forgetting problem where the fine-tuned LLMs will lose pre-trained knowledge. Therefore, in addition to the performance degradation of the LLMs on the original domain, risks arise that attackers can utilize fine-tuned datasets to break the defense techniques in the pre-trained LLMs~(e.g., adversarial-trained LLMs) and find backdoors to attack the LLM systems.  

\ul{\textit{Risk8: Risks in the model hub}}. In the LLM fine-tuning process, developers usually download pre-trained LLMs from the open-source model hub such as Hugging Face~\cite{huggingface}. Even though such model hubs have a security scan procedure to check the reliability of uploaded LLMs, current work~\cite{zhao2024models} demonstrated that these scanning techniques can be easily bypassed. Therefore, the download LLMs might contain different security issues and need to be carefully checked before the fine-tuning process.

No matter which type of deep learning model, deploying it to real-world applications is the final goal. A recent report~\cite{airisk} shows that most safety issues~(24\%) happen at the system level. Existing works pay more attention to the model-level evaluation for security assurance but lack of study of LLM-based applications. Since the downstream participants of LLM-based applications are users, the security risks will cause practical damage. 

\ul{\textit{Risk9: Risks introduced by model optimization}}. Recent works~\cite{egashira2024exploiting} revealed that attackers can utilize model compression techniques to hide backdoors to the LLM where the original LLM is secure but the compressed LLM is vulnerable. Therefore, the risks can be introduced by the model optimization process, and it is necessary to carefully choose the model compression techniques and avoid potential attacks.

\ul{\textit{Risk10: Vulnerability in other software components}}. In addition to the model, LLM applications could also contain other software components. For example, a LLM-driven autonomous driving system could contain multiple components such as the perception module, the planning module, and the control module. Software components could affect each other and influence the final output of the system, e.g., the unreliable images recorded by the camera could mislead the decision of the following components in the LLM system. Therefore, it is necessary to analyze the influence and dependencies of such software components in the LLM application security.

\ul{\textit{Risk11: Security risks via user feedback}}. LLM providers will also utilize user feedback (prompts and answers) to update and maintain models for domain adaptation. In this situation, the LLM provider could be the downstream participant while users are the upstream suppliers. Unfortunately, there is a concern that the user feedback could also introduce security-critical samples to the fine-tuning dataset which can be exploited by attackers, and then harm the reliability of LLMs.

\ul{\textit{Risk12: Risks due to the unknown tasks and data distribution}}. Every evaluation is based on a prepared test dataset that covers a certain distribution domain, however, it is unable to estimate the usage environment of the LLM-based applications. Thus, it is impossible to guarantee that the pre-defined risk assessments can always work. The security issues of the real-world LLM applications are the most difficult to handle.

\section{Risk Mitigation Discussion}

Based on our analysis, we provide guidelines for future research on secure LLM SC from data, model, and application three perspectives. 


\ul{\textit{Security assurance in the data construction.}} First, for data collectors, be careful to use automatic data selection and data cleaning methods to collect datasets as they can be easily fooled. Then, the upstream supplier is recommended to provide some data distribution information to the downstream users such as the top-N token frequency distribution. In this way, developers have chances to control the domain distribution in the following model testing and fine-tuning processes to report more precise and reliable evaluation results, and prepare fine-tuned models without harming the original distribution.

\ul{\textit{Security assurance in the model preparation.}} LLM preparation is the key component in the LLM SC. In the training phase, fully-supervised active learning should be carefully employed as it could reduce the robustness of the model and can be attacked. Semi-supervised active learning with data augmentation is a better choice that can mitigate the attacks with data modification. Considering the downstream action model deployment of LLMs, developers need to think about whether it is necessary to use quantization-aware~(or other compression-aware) training techniques. Besides, it is useful to highlight data samples that are uncertain to the LLM before compression as the compressed LLM could produce different outputs as shown in previous works~\cite{xie2019diffchaser}.

\ul{\textit{Security assurance in the application development}} The upstream participant of applications is user. Thus, security assurance of LLM-based applications is a crucial and emerging task. First, every LLM-based application needs a test report on the targeted domains that covers diverse evaluation metrics~(especially security risk evaluation) to guide its reliable usage. Besides, as mentioned in Section~\ref{sec:llm_sc_security_analysis}, multiple types of attacks can be injected into applications and some can bypass existing defense techniques such as ModelScan. Therefore, application maintainers must design stronger scan techniques to check uploaded applications, specifically focusing on embedded operators~(APIs) that can assess users' devices. 

Quality assurance of a single component in the LLM SC is not enough to ensure the reliability of the final produced LLM systems. In the real-world LLM SC,  upstream suppliers can directly affect its connected downstream participants and indirectly related participants in different aspects. It is essential to design metrics and criteria to measure the influence from one component to another for risk assurance of the whole SC.  

\section{Future Plan}

In the future, we plan to:

\begin{itemize}[leftmargin=*]

\item extend this work and make a comprehensive survey on LLM SC security risks.

\item design metrics and criteria to measure the influence caused by security issues between different components in the LLM SC. 

\item propose techniques for security assurance of LLM SC. For example, we plan to design multi-objective guided data selection methods for targeted instruction tuning~\cite{xia2024less} to efficiently train LLMs with better performance while avoiding potential attacks on the data selection process. 

\end{itemize}

\section{Conclusion}
\label{sec: conclusion}

In this paper, we have initiated a discussion about security risks in the large language model supply chain. Different from existing works that mainly focused on studying a single component~(especially the model) in the SC, we explored the potential risks that lie in the integration of the SC from upstream suppliers to downstream users. In total, we summarized 12 risks that gained limited attention by current research and provided promising guidance for both researchers and developers to help construct more reliable LLM systems. We believe ensuring the security of LLM SC is crucial and our work can facilitate the understanding of the risks.

\bibliographystyle{IEEEtran}
\balance
\bibliography{IEEEabrv,main}

\end{document}